\documentstyle[amssymb,epsf,cite,11pt]{article}

\newcommand{\be}{\begin{equation}}
\newcommand{\ee}{\end{equation}}

\begin{document}

\title{\bf \Large Determination of multifractal dimensions of complex networks by means of the sandbox algorithm}

\author{  Jin-Long Liu$^{1}$, Zu-Guo Yu$^{1,2}$\thanks{
  Corresponding author, email: yuzg1970@yahoo.com}, and Vo Anh$^{2}$ \\
{\small $^{1}$Hunan Key Laboratory for Computation and Simulation in Science and Engineering and Key}\\
{\small Laboratory of Intelligent Computing and Information Processing of Ministry of Education,}\\
{\small Xiangtan University, Xiangtan,  Hunan 411105, China.}\\
{\small $^{2}$School of Mathematical Sciences, Queensland University of Technology, GPO Box 2434,}\\
{\small Brisbane, Q4001, Australia.} }
\date{}
\maketitle

\begin{abstract}

Complex networks have attracted much attention in diverse areas of
science and technology. Multifractal analysis (MFA) is a useful
way to systematically describe the spatial heterogeneity of both
theoretical and experimental fractal patterns. In this paper,
we employ the sandbox (SB) algorithm proposed by
T\'{e}l {\it et al.} ({\it Physica A}, 159 (1989) 155-166), for
MFA of complex networks. First we compare the SB algorithm with
two existing algorithms of MFA for complex networks: the
compact-box-burning (CBB) algorithm proposed by Furuya and Yakubo
({\it Phys. Rev. E}, 84 (2011) 036118), and the improved
box-counting (BC) algorithm proposed by Li {\it et al.} ({\it J.
Stat. Mech.: Theor. Exp.}, 2014 (2014) P02020) by calculating the
mass exponents $\tau(q)$ of some deterministic model networks. We
make a detailed comparison between the numerical and theoretical
results of these model networks. The comparison results show that
the SB algorithm is the most effective and feasible algorithm to
calculate the mass exponents $\tau(q)$ and to explore the
multifractal behavior of complex networks. Then we apply the SB
algorithm to study the multifractal property of some classic model
networks, such as scale-free networks, small-world networks, and
random networks. Our results show that multifractality exists in
scale-free networks, that of small-world networks is not obvious,
and it almost does not exist in random networks.

\end{abstract}


\section{Introduction}

Many studies have shown that complex networks play an
important role in characterizing complicated dynamic systems in
nature and society \cite{Song2005}. This is because the nodes of a
complex network represent the elements, and the edges represent
and simplify the complexity of their interactions so that we can
better focus on the topological relation between two elements in a
complex system. Recently, complex networks have attracted the
attention of a lot of researchers from different fields. Based on
the self-similarity of fractal geometry
\cite{Mandelbrot1983, Feder1988, Falconer1997}, Song \textit{et
al.} \cite{Song2005} generalized the box-counting algorithm and
used it in the field of complex networks. They found that many
complex networks are self-similar under certain length-scales. The
fractal dimension has been widely used to characterize complex
fractal sets \cite{Mandelbrot1983, Feder1988, Falconer1997}.
Because the metric on graphs is not the same as the Euclidian
metric on Euclidian spaces, the box-counting algorithms to
calculate the fractal dimension of a network is much more
complicated than the traditional box-counting algorithm for
fractal sets in Euclidian spaces. Song \textit{et al.}
\cite{Song2007} developed some algorithms to calculate the fractal
dimension of complex networks. Then Kim \textit{et al.}
\cite{Kim2007, Kim2007s} proposed an improved algorithm to investigate the
skeleton of networks and the fractal scaling in scale-free networks. Zhou \textit{et al.} \cite{ZhouJiang2007}
proposed an algorithm based on the edge-covering box counting to
explore self-similarity of complex cellular networks. Later on, a
ball-covering approach \cite{GaoHu2008} and an approach defined by
the scaling property of the volume \cite{GuoCai2009, Shanker2007}
were proposed for fractal dimensions of complex networks. The features of topology and statistics \cite{Gao2009, Bianconi2007}, the
fractality and percolation transition \cite{Rozenfeld2009},
fractal transition \cite{Rozenfeld2010} in complex networks, and
properties of a scale-free Koch networks \cite{Zhang2008,
Zhang2009, Liu2010, Zhang2010, Dai2013} have turned out to be hot topics in recent
years.

As a generalization of fractal analysis, the tool of multifractal
analysis (MFA) may have a better performance on characterizing the
complexity of complex networks in real world. MFA has been widely
applied in a variety of fields such as financial modelling
\cite{Canessa2000, Anh2000}, biological systems \cite{Yu2001a, Yu2001b, Yu2003, Yu2004, Yu2006, Yu2010b, Anh2002,
Zhou2005, Han2010, Zhu2011}, and geophysical data analysis \cite{SchertzerL1987, Lovejoy1996, Lovejoy2010a,
Lovejoy2010b, Kantelhardt2006, Veneziano2006, Venugopal2006,
Wanliss2005, Yu2009, Yu2010a}.

In recent years, MFA also has been successfully used in complex
networks and seems more powerful than fractal analysis. Lee and
Jung \cite{Lee2006} found that MFA is the best tool to describe
the probability distribution of the clustering coefficient of a
complex network. Some algorithms have been proposed to calculate
the mass exponents $\tau(q)$ and to study the multifractal
properties of complex networks \cite{Furuya2011, Wang2012, Li2014,
Liu2014}. Based on the compact-box-burning algorithm for fractal
analysis of complex networks which is introduced by Song {\it et
al.} \cite{Song2007}, Furuya and Yakubo \cite{Furuya2011} proposed
a {\it compact-box-burning} (CBB) algorithm for MFA of complex
networks and applied it to show that some networks have
multifractal structures. Wang {\it et al.} \cite{Wang2012}
proposed a modified fixed-size box-counting algorithm to detect
the multifractal behavior of some theoretical and real networks.
Li {\it et al.} \cite{Li2014} improved the
 modified fixed-size box-counting algorithm \cite{Wang2012} and
used it to investigate the multifractal properties of a family of
fractal networks introduced by Gallos {\it et al.}
\cite{Gallos2007}. We call the algorithm in Ref. \cite{Li2014} the
{\it improved BC} algorithm. Recently, we adopted the improved BC
to study the multifractal properties of the recurrence networks
constructed from fractional Brownian motions \cite{Liu2014}.

In order to easily obtain the generalized fractal dimensions of
real data, T\'{e}l {\it et al.} \cite{Tel1989} introduced a
sandbox algorithm which is originated from the box-counting
algorithm \cite{Halsey1986}. They \cite{Tel1989} pointed out that
the sandbox algorithm gives a better estimation of the generalized
fractal dimensions in practical applications. So far, the sandbox
algorithm also has been widely applied in many fields. For
example, Yu {\it et al.} \cite{Yu2004} used it to perform MFA on
the measures based on the chaos game representation of protein
sequences from complete genomes.

In this article, we employ the sandbox (SB) algorithm
proposed by T\'{e}l {\it et al.} \cite{Tel1989} for MFA of complex
networks. First we compare the SB algorithm with the CBB and
improved BC algorithms for MFA of complex networks in detail by
calculating the mass exponents $\tau(q)$ of some deterministic
model networks. We make a detailed comparison between the
numerical and theoretical results of these model networks. Then we
apply the SB algorithm to study the multifractal property of some
classic model networks, such as scale-free networks, small-world
networks, and random networks.

\section{Sandbox algorithm for multifractal analysis of complex networks}

It is well known that the fixed-size box-covering
algorithm \cite{Halsey1986} is one of the most common and
important algorithms for multifractal analysis. For a given
measures $\mu$ with support set $E$ in a metric space, we consider
the following partition sum
\begin{equation}
Z_{\epsilon}(q) = \sum_{\mu(B)\neq 0}[\mu(B)]^{q},
\end{equation}
$q \in R$, where the sum runs over all different nonempty boxes
$B$ of a given size $\epsilon$ in a box covering of the support
set $E$. From the definition above, we can easily obtain
$Z_{\epsilon}(q) \geq 0$ and $Z_{\epsilon}(0) = 1$. The mass
exponents $\tau(q)$ of the measure $\mu$ can be defined as
\begin{equation}
\tau(q) = \lim_{\epsilon\rightarrow 0}\frac{\ln Z_{\epsilon}(q)}{\ln \epsilon}.
\end{equation}
Then the generalized fractal dimensions $D(q)$ of the measure $\mu$ are defined as
\begin{equation}
D_{q} = \frac{\tau(q)}{q-1}, ~~~~~~\textrm{for}~~ q \neq 1,
\end{equation}
and
\begin{equation}
D_{q} = \lim_{\epsilon\rightarrow0}\frac{Z_{1,\epsilon}}{\ln \epsilon}, ~~~\textrm{for}~~ q = 1,
\end{equation}
where $Z_{1,\epsilon} = \sum_{\mu(B)\neq0} \mu(B)\ln\mu(B)$. The
linear regression of $[\ln Z_{\epsilon}(q)]/(q-1)$ against $\ln
\epsilon$ for $q \neq 1$ gives a numerical estimation of the
generalized fractal dimensions $D_{q}$, and similarly a linear
regression of $Z_{1,\epsilon}$ against $\ln \epsilon$ for $q = 1$.
In particular, $D_{0}$ is the box-counting dimension (or fractal
dimension), $D_{1}$ is the information dimension, and $D_{2}$ is
the correlation dimension.

In complex network, the measure $\mu$ of each box can be defined
as the ratio of the number of nodes covered by the box and the
total number of nodes in the entire network. In addition, we can
determine the multifractality of complex network by the shape of
$\tau(q)$ or $D(q)$ curve. If $D(q)$ is a constant or $\tau(q)$ is
a straight line, the object is monofractal; on the other hand, if
$D(q)$ or $\tau(q)$ is convex, the object is multifractal.

Before we use the following SB algorithm to perform MFA of a
network, we need to apply the Floyd's algorithm \cite{Floyd1962}
of Matlab-BGL toolbox \cite{Gleich} to calculate the shortest-path
distance matrix $D$ of this network according to its adjacency
matrix $A$.

The sandbox algorithm proposed by T\'{e}l {\it et al.}
\cite{Tel1989} is an extension of the box-counting algorithm
\cite{Halsey1986}. The main idea of this sandbox algorithm is that
we can randomly select a point on the fractal object as the center
of a sandbox and then count the number of points in the sandbox.
The generalized fractal dimensions $D(q)$ are defined as
\begin{equation}
D_{q} = \lim_{r \rightarrow 0}\frac{\ln \langle[M(r)/M(0)]^{q-1}\rangle}{\ln(r/d)}\frac{1}{q-1},~~~~q \in R,
\end{equation}
where $M(r)$ is the number of points in a sandbox with a radius of $r$, $M(0)$ is the total number of points in the fractal
object. The brackets $\langle \cdot \rangle$ mean to take statistical average over randomly chosen centers of the sandboxes. As
a matter of fact, the above equation can be rewritten as
\begin{equation}
\ln(\langle [M(r)]^{q-1} \rangle)\ \propto \
D(q)(q-1)\ln(r/d)+(q-1)\ln(M_{0}).
\end{equation}
So, in practice, we often estimate numerically the generalized
fractal dimensions $D(q)$ by performing a linear regression of
$\ln(\langle [M(r)]^{q-1} \rangle )$ against $(q-1)\ln(r/d)$; and
estimate numerically the mass exponents $\tau(q)$ by performing a
linear regression of $\ln(\langle [M(r)]^{q-1} \rangle )$ against
$\ln(r/d)$. In a complex network, we can randomly choose a node of
a network as the center of a sandbox. $M(r)$ and $M(0)$ represent
the number of nodes in the sandbox of radius $r$ and the size of
the network, respectively. The SB algorithm for MFA of complex
networks can be described as follows.

\begin{enumerate}

\item[(i)]  Initially, make sure all nodes in the entire network are not selected as a center of a sandbox.

\item[(ii)] Set the radius $r$ of the sandbox which will be used to cover the nodes in the range $r \in [1, d]$,
where $d$ is the diameter of the network.

\item[(iii)] Rearrange the nodes of the entire network into
random order. More specifically, in a random order, nodes which
will be selected as the center of a sandbox are randomly arrayed.

\item[(iv)] According to the size $N$ of networks, choose the
first 1000 nodes in a random order as the center of 1000
sandboxes, then search all the neighbor nodes by radius $r$ from
the center of each sandbox.

\item[(v)] Count the number of nodes in each sandbox of radius $r$, denote the number of nodes in each sandbox as $M(r)$.

\item[(vi)] Calculate the statistical average $\langle [M(r)]^{q-1} \rangle$ of $[M(r)]^{q-1}$ over all 1000 sandboxes of radius $r$.

\item[(vii)] For different values of $r$, repeat steps (ii) to (vi) to
calculate the statistical average $\langle [M(r)]^{q-1} \rangle$
and then use $\langle [M(r)]^{q-1} \rangle$ for linear regression.
\end{enumerate}

We need to choose an appropriate range of $r \in [r_{min},
r_{max}]$, then calculate the generalized fractal dimensions
$D(q)$ and the mass exponents $\tau(q)$ in this scaling range. In
our calculation, we perform a linear regression of $\ln(\langle
[M(r)]^{q-1} \rangle)$ against $\ln(r/d)$  and then choose the
slope as an approximation of the mass exponents $\tau(q)$ (the
process for estimating the generalized fractal dimensions $D(q)$ is
similar).

For the improved BC and CBB algorithms, we need to cover the
entire network by repeating a large number of same steps and then
to find the minimum possible number of boxes by performing many
realizations. Then we can choose an appropriate range of $r \in
[r_{min}, r_{max}]$ to calculate the mass exponents $\tau(q)$ by
performing a linear regression of $\ln Z_{r/d}(q)$ against
$\ln(r/d)$. In the process of finding a covering of the network,
the two existing algorithms require that each node in the network
cannot be covered by more than one box at the same time. In
addition, for the CBB algorithm, we have to randomly select many
nodes $p$ from the candidate set $C$ to form a compact box and
then repeat these steps until the entire network is covered.
Because of these limitations, these two algorithms must take a
large amount of CPU time and memory resources to record
information of the nodes which have been covered in the previous
steps. From the above descriptions of SB algorithm, however, we
find that a big difference from the improved BC and CBB algorithms
is that SB algorithm only requires to randomly choose some nodes
as the center of a sandbox and then to count the number of nodes
in each sandbox, hence we don't need so many sandboxes to cover
the entire network. Moreover, we focus on the number of nodes of
each sandbox, so we also don't need to know whether or not the
nodes in the sandbox have been covered by other sandboxes. For the
SB algorithm, therefore, we only consume a very small amount of
CPU time and memory resources. In this sense, the SB algorithm can
be considered to be the most effective and feasible algorithm for
MFA of complex networks.

\section{Algorithm comparison}

Now we compare the SB algorithm with the CBB and improved
BC algorithms for MFA of complex networks \cite{Furuya2011,Li2014}
in detail by calculating the mass exponents $\tau(q)$ of some
deterministic model networks. We make a detailed comparison
between the numerical and theoretical results of these model
networks.

\subsection{The model networks}

In the past two decades, many network models have been
introduced to study and simulate the topological and fractal
properties and the growth mechanism of many complex dynamical
systems in real world. Watts and Strogatz \cite{Watts1998}, Newman
and Watts \cite{Newman1999} proposed the WS and NW small-world
network models to explain the small-world character of many real
complex networks respectively. In order to reveal the generating
principle of power law distributions, Barab\'{a}si {\it et al.}
\cite{Barabasi1999} proposed a scale-free network model (BA model)
based on the growth and preferential attachment characteristics of
real networks.

In 2002, Dorogovtsev {\it et al.} \cite{Dorogovtsev2002}
introduced the simple deterministic graphs to model scale-free
networks. They pointed out that the family of deterministic
networks (DGM networks) are pseudo-fractals.  In order to
understand the self-similarity of complex networks better,
Rozenfeld {\it et al.} \cite{Rozenfeld2007} generalized the DGM
network to a family of scale-free networks, namely $(u,
v)$-flowers and $(u, v)$-trees. The DGM network is a special case
of the $(1, v)$-flowers. For $u = 1$, the networks are
self-similar only in the weak sense. For $v \geq u > 1$, the
networks are self-similar and possess the well-defined fractal
dimension. In 2006, Song {\it et al.} \cite{Song2006} proposed the
minimal model to study the evolved law of complex networks and
then simulate the emergence of self-similarity and small-world
properties of these networks. Later on, Gallos {\it et al.}
\cite{Gallos2007} introduced a generalized version of the minimal
model proposed by Song {\it et al.} \cite{Song2006}.

Based on the above network models, some researchers have studied
analytically and numerically the fractal and multifractal
properties of networks generated from these models. Song {\it et
al.} \cite{Song2006} and Gallos {\it et al.} \cite{Gallos2007}
gave an analytical formula to calculate the fractal dimensions of
the minimal model and its generalized version respectively. Then
Rozenfeld {\it et al.} \cite{Rozenfeld2007} also put forward a
theoretical framework for computing the degree exponent and the
fractal dimension of the $(u, v)$-flower network. In addition,
Furuya and Yakubo \cite{Furuya2011} analytically and numerically
investigated the multifractal properties of several deterministic,
stochastic, and real-world fractal scale-free networks. They gave
an analytical formula of the mass exponents $\tau(q)$ for some
class of fractal scale-free model networks by a mean-field
approximation. They define the mass exponents $\tau(q)$ of a
complex network as
\begin{equation}
\tau(q)=
\left\{
\begin{array}{cc}
(q-1)D_{f},~~~~~q < \gamma -1,\\
qD_{f}\frac{\gamma - 2}{\gamma - 1},~~~~~~~q \geq \gamma - 1,
\end{array}\right.
\end{equation}
where $D_{f}$ is the fractal dimension, $\gamma$ is the degree
exponent of the complex network.

For the $(u, v)$-flower, we start with a cycle graph consisting of
$u+v$ nodes and $u+v$ links. Then we can obtain the $(u,
v)$-flower network of generation $n+1$ by replacing each link in
$n$th generation network by two parallel paths with length (number
of links of the path) $u$ and $v$ respectively. In Fig. 1, we show
how to construct the $(2, 2)$-flower network as an example.
Rozenfeld {\it et al.} \cite{Rozenfeld2007} gave the analytical
formulas of degree exponent $\gamma$ and fractal dimension $D_{f}$
of $(u, v)$-flower. The degree exponent $\gamma$ of the $n$th
generation $(u, v)$-flower network is given by
\begin{equation}
\gamma = 1+ \frac{\ln (u+v)}{\ln 2},
\end{equation}
and the fractal dimension $D_{f}$ is presented by
\begin{equation}
D_{f}=\frac{\ln (u+v)}{\ln u},  ~~u>1, v\ge u.
\end{equation}
By substituting Eqs. (8) and (9) into Eq. (7), the mass exponents
$\tau(q)$ of the $(u, v)$-flower network can be written as
\begin{equation}
\tau(q)=
\left\{
\begin{array}{cc}
(q-1)\frac{\ln (u+v)}{\ln u},~~~~~q < \frac{\ln (u+v)}{\ln 2},\\
q\frac{\ln ((u+v)/2)}{\ln u},~~~~~~~~q \geq \frac{\ln (u+v)}{\ln 2}.
\end{array}\right.
\end{equation}

\begin{figure}
\centerline{\epsfxsize=8.5cm \epsfbox{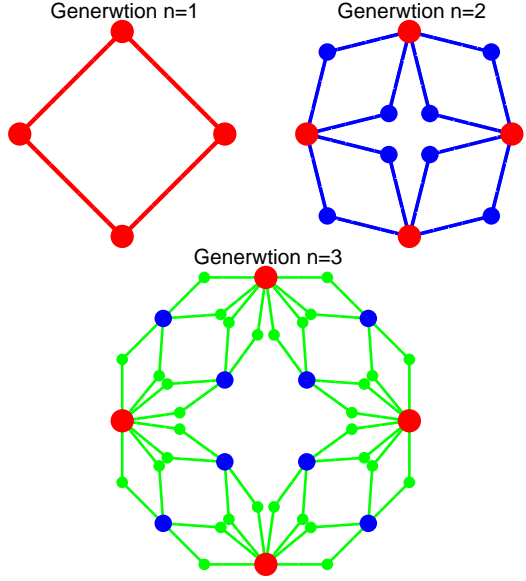}}
\caption{Construction of a $(u, v)$-flower network. Example of the $(u, v)$-flower network model of generations $n=1, 2, 3$ with parameters $u=2$ and $v=2$.}\label{1}
\end{figure}

The minimal model proposed by Song {\it et al.} \cite{Song2006} is
a probabilistic combination of two different connectivity modes:
Mode I with probability $e$ and Mode II with probability $1-e$.
Mode I means that all the old connections generated in the
previous generation are remained; Mode II means that we remove all
the old connections generated in the previous generation and add a
new edge to connect two new generated nodes. Before using Modes I
and II, we attach $m$ new nodes to each endpoint of each edge $l$
in the network of the current generation. A remarkable advantage
of this stochastic combination of the two different growth modes
is that its level of fractality can be controlled by the
probability $e$. Song et al. \cite{Song2006} pointed out that the
minimal model is a pure fractal network when the probability
$e=0$, and a pure small-world network when $e=1$. In this paper,
we only consider the minimal model with probability $e=0$. We
start with a star structure as in Ref. \cite{Song2006}. Then
we obtain the minimal model of generation $n+1$ by adding $mk_{i}$
new nodes to each node $i$ with degree $k_{i}$ of generation $n$,
where $m$ is a given parameter. In addition, we adopt the growth
Mode II to replace all the old connections in the previous
generation. In Fig. 2, we show how to construct the pure fractal
network with parameters $m=2$ and $e=0$ as an example. As pointed
out by Song {\it et al.} \cite{Song2006}, the degree exponent
$\gamma$ is
\begin{equation}
\gamma = 1+ \frac{\ln (2m+1)}{\ln m},
\end{equation}
and the fractal dimension $D_{f}$ is
\begin{equation}
D_{f}=\frac{\ln (2m+1)}{\ln 3}.
\end{equation}
By substituting Eqs. (11) and (12) into Eq. (7), the mass
exponents $\tau(q)$ of this minimal model can be written as
\begin{equation}
\tau(q)=
\left\{
\begin{array}{cc}
(q-1)\frac{\ln (2m+1)}{\ln 3},~~~~~q < \frac{\ln (2m+1)}{\ln m},\\
q\frac{\ln ((2m+1)/m)}{\ln 3},~~~~~~~q \geq \frac{\ln (2m+1)}{\ln m}.
\end{array}\right.
\end{equation}

\begin{figure}
\centerline{\epsfxsize=8.5cm \epsfbox{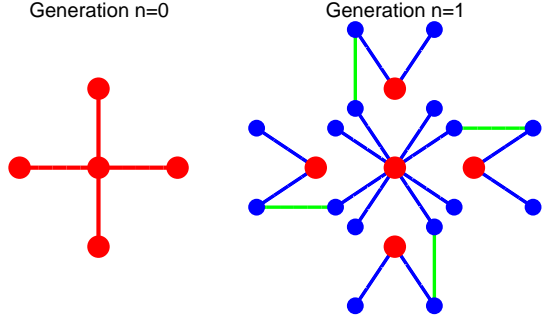}}
\caption{Construction of a pure fractal network. Example of the minimal model network of generations $n=0, 1$ with parameters $m=2$ and $e=0$.}\label{2}
\end{figure}

Based on the minimal model, Gallos {\it et al.} \cite{Gallos2007}
proposed a generalized version of this model. Here we start with
two nodes and one edge between them as in Ref.
\cite{Gallos2007}. Then we obtain the network of next
generation by attaching $m$ new nodes to each endpoint of each
edge $l$ in the network of the current generation. With
probability $e$, each edge $l$ of the current generation is
remained and $x-1$ $(x \leq m)$ new edges are added to connect
pairs of new nodes attached to the endpoints of $l$. Otherwise, we
remove each edge $l$ in the network of the current generation and
add $x$ $(x \leq m)$ new edges to connect pairs of nodes attached
to the endpoints of $l$. In this paper, we only consider the
network with probability $e=0$. In Fig. 3, we show how to
construct the generalized network with parameters $m=2$, $x=2$,
and $e=0$, as an example. In the case of probability $e=0$, from
Refs. \cite{Li2014, Rozenfeld2009}, the degree exponent
$\gamma$ is
\begin{equation}
\gamma = 1+ \frac{\ln (2m+x)}{\ln m},
\end{equation}
and the fractal dimension $D_{f}$ is
\begin{equation}
D_{f}=\frac{\ln (2m+x)}{\ln 3}.
\end{equation}
By substituting Eqs. (14) and (15) into Eq. (7), we can obtain the
mass exponents $\tau(q)$ of the generalized version of the minimal
model, which can be written as
\begin{equation}
\tau(q)=
\left\{
\begin{array}{cc}
(q-1)\frac{\ln (2m+x)}{\ln 3},~~~~~q < \frac{\ln (2m+x)}{\ln m},\\
q\frac{\ln ((2m+x)/m)}{\ln 3},~~~~~~~q \geq \frac{\ln (2m+x)}{\ln m}.
\end{array}\right.
\end{equation}

\begin{figure}
\centerline{\epsfxsize=8.5cm \epsfbox{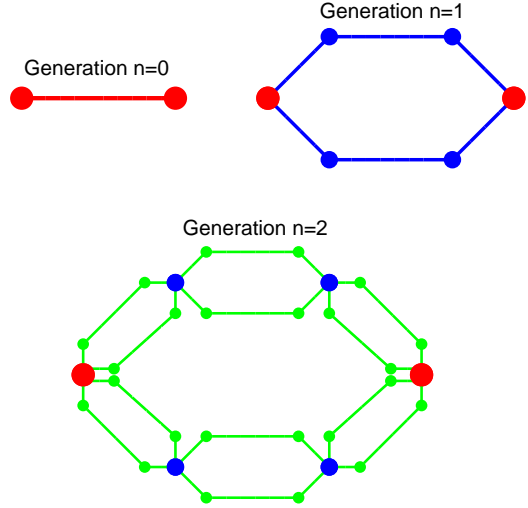}}
\caption{Construction of a generalized network of the minimal model. Example of the generalized network of generations $n=0, 1, 2$ with parameters $m=2$, $x=2$, and $e=0$.}\label{3}
\end{figure}

In the following, we generate networks using the three models and
numerically study their multifractality by the SB, CBB and
improved BC algorithms (in Section II). Then we give a detailed
comparison between the three algorithms based on the numerical
results and theoretical formulas of the mass exponents $\tau(q)$
for these model networks.

\subsection{Comparison results}

In this work, we set the range of the $q$ values from $-10$ to
$10$ with a step of $1/3$. In order to compare with the results in
Ref. \cite{Furuya2011}, we generated the $(u, v)$-flower
network with $u = 2$ and $v = 2$. Considering the limitation of
the computational capacity of our computer, we only constructed
the 7th generation $(u, v)$-flower network. An important step of
calculating the mass exponents $\tau(q)$ and the generalized
fractal dimensions $D(q)$ is to obtain the appropriate range of
$r$ (i.e. $r \in [r_{min},r_{max}]$). As an example, we show the linear
regressions of the $\ln(\langle [M(r)]^{q-1} \rangle)$ vs $\ln
(r/d)$ in the SB algorithm for the 7th generation $(u, v)$-flower
network with $u = 2$ and $v = 2$ in Fig. 4. We selected
the range of $r$ as $[2,20]$ to fit these data points. From Fig. 4, we can
observe the apparent power law behaviors for the 7th generation
$(u, v)$-flower network with $u = 2$ and $v = 2$. So we selected
the linear fit scaling range of $r$ as $[2,20]$ to calculate the
mass exponents $\tau(q)$. In Fig. 5, we show the mass exponents
$\tau(q)$ of the $(u, v)$-flower network calculated by the SB, CBB
and improved BC algorithms. From Fig. 5, we can see that the
numerical results obtained by the three algorithms are consistent
with the theoretical results.

For the minimal model network, we started with a star structure of
5 nodes as in Ref. \cite{Song2006} and then generated the 5th
generation network with $m = 2$ and $e = 0$. We calculated the
mass exponents $\tau(q)$ by the three algorithms. The numerical
results are shown in Fig. 6. From Fig. 6, we can see that the
numerical results obtained by the three algorithms agree well with
the theoretical results.

In addition, we also generated the generalized version of the
minimal model with $m = 2$, $x = 2$, and $e = 0$. Here we only
constructed the 5th generation of the generalized network. The
numerical results are shown in Fig. 7. From Fig. 7, we can see
that the numerical results obtained by the three algorithms have
good agreement with the theoretical results.

\begin{figure}
\centerline{\epsfxsize=8.5cm \epsfbox{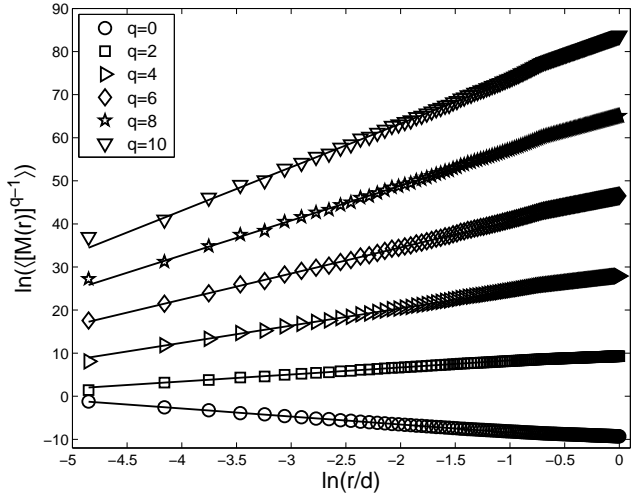}}
\caption{Linear regressions for calculating the mass exponents $\tau(q)$ of the 7th generation $(u, v)$-flower network with $u = 2$ and $v = 2$. The result is calculated by the SB algorithm.}\label{4}
\end{figure}

\begin{figure}
\centerline{\epsfxsize=8.5cm \epsfbox{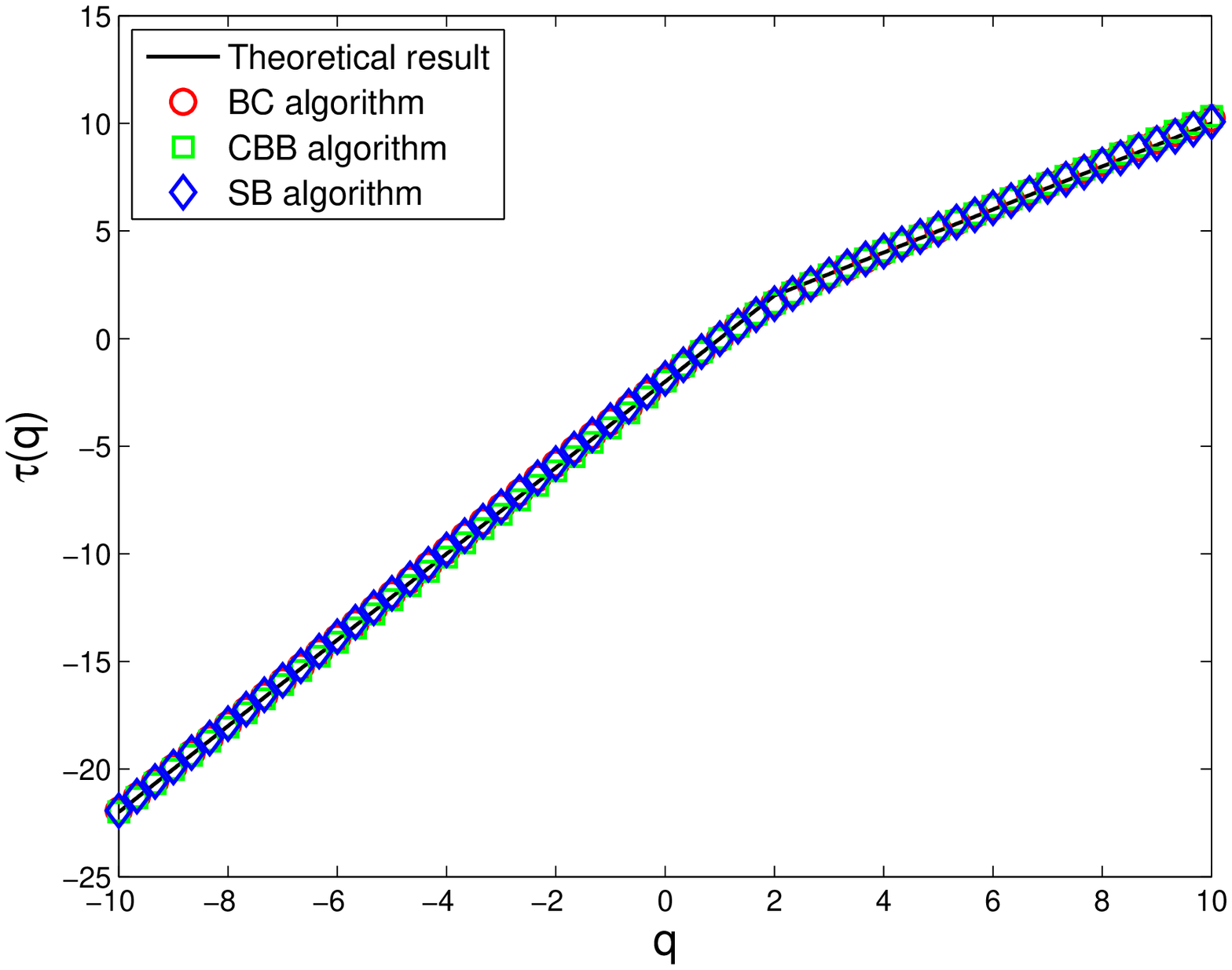}}
\caption{(Color online) Mass exponents $\tau(q)$ for the 7th generation $(u, v)$-flower network with $u = 2$ and $v = 2$. Solid
line (black line) represents the mass exponents $\tau(q)$ given by Eq. (10). The curves indicated by symbols (circles, squares,
and diamonds) represent the numerical estimation of the mass exponent $\tau(q)$ calculated by the three algorithms for MFA of
complex networks, respectively.}\label{5}
\end{figure}

\begin{figure}
\centerline{\epsfxsize=8.5cm \epsfbox{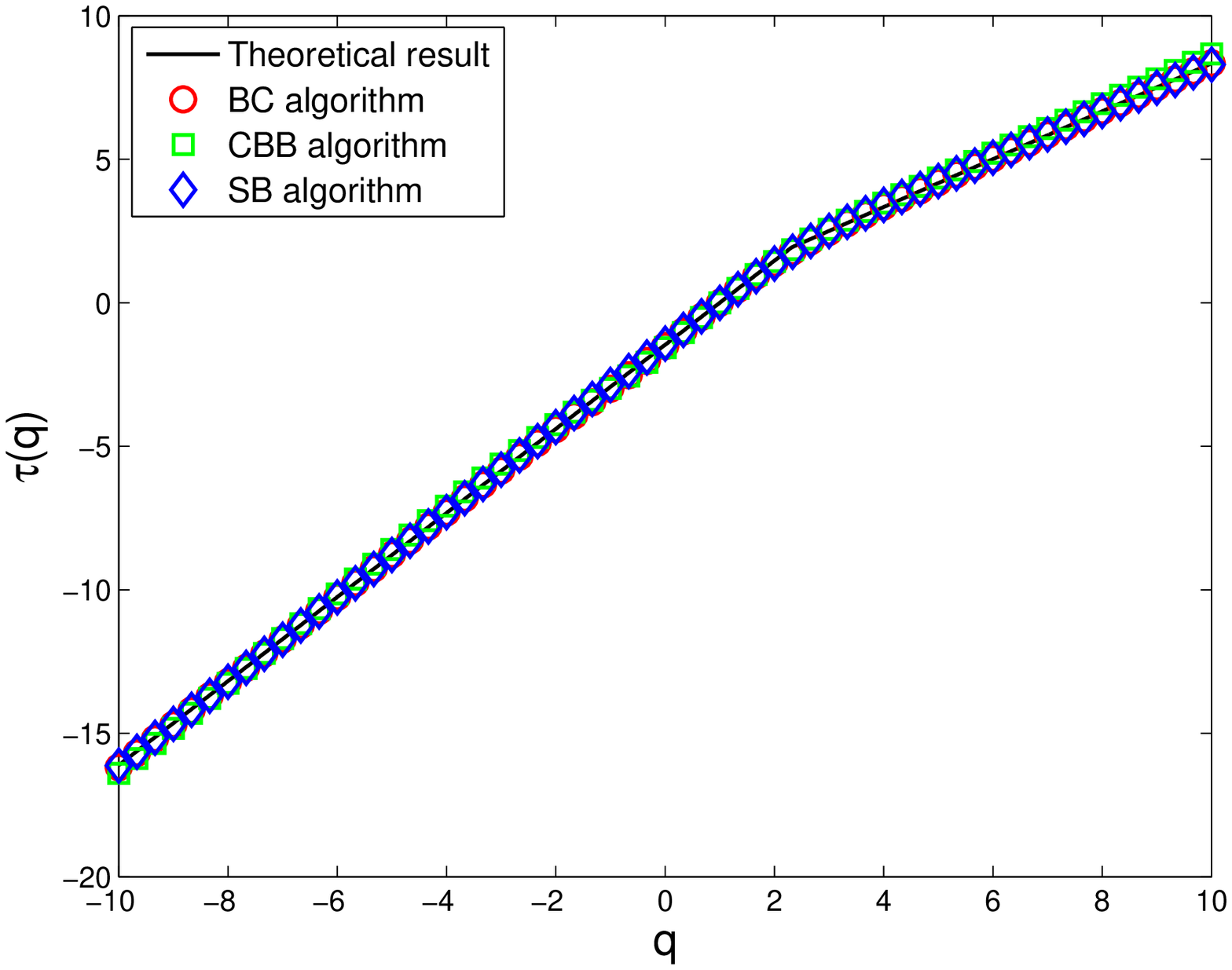}}
\caption{(Color online) Mass exponents $\tau(q)$ for the 5th
generation minimal model network with $m = 2$ and $e = 0$. Solid line
(black line) represents the mass exponent $\tau(q)$ given by Eq. (13). The curves indicated by symbols (circles, squares, and
diamonds) represent the numerical estimation of the mass exponent $\tau(q)$ calculated by the three algorithms for MFA of
complex networks, respectively.}\label{6}
\end{figure}

\begin{figure}
\centerline{\epsfxsize=8.5cm \epsfbox{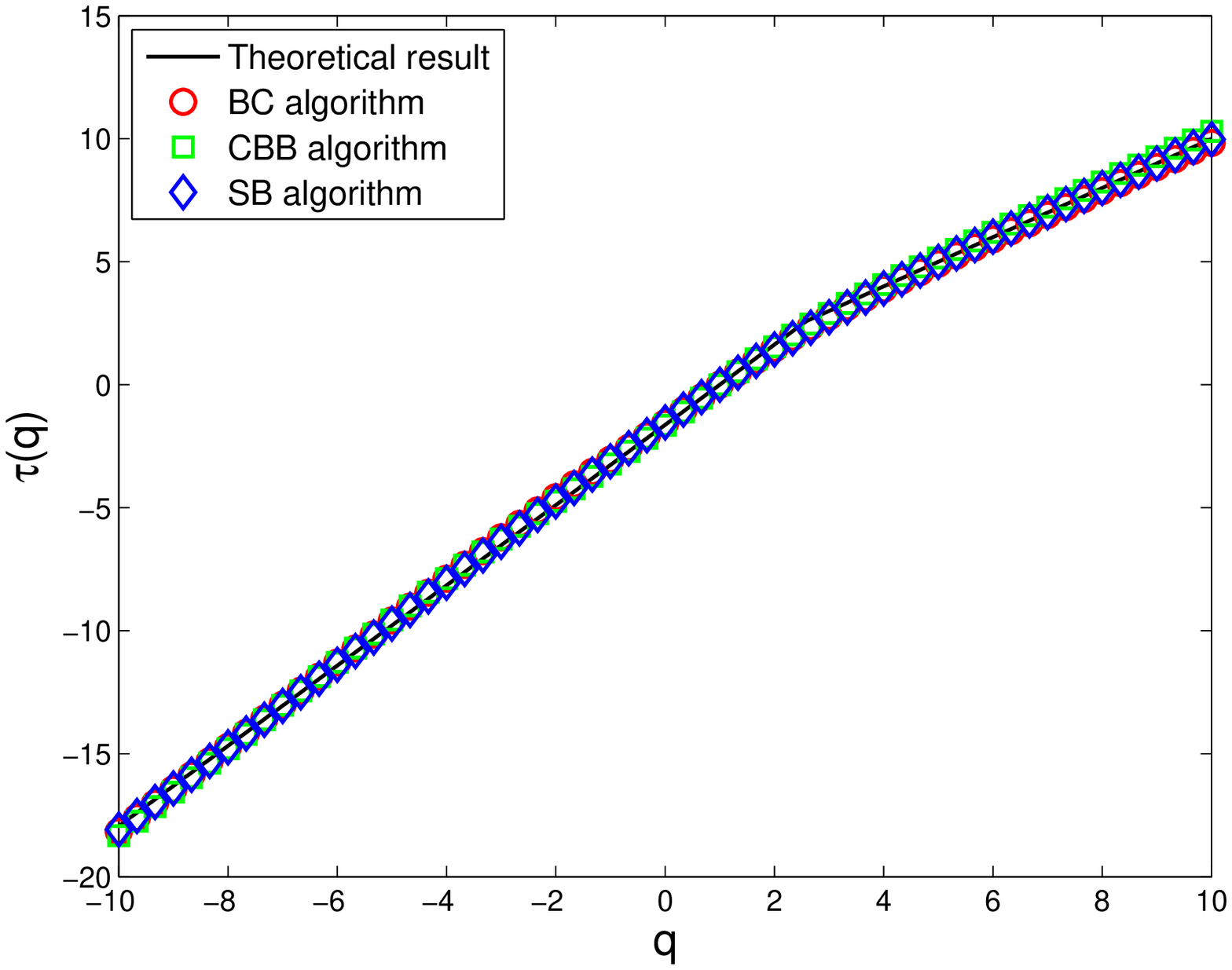}}
\caption{(Color online) Mass exponents $\tau(q)$ for the 5th generation network from generalized version of the minimal model with $m = 2$, $x = 2$,
and $e = 0$. Solid line (black line) represents the mass exponent $\tau(q)$ given by Eq. (16). The curves indicated by symbols
(circles, squares, and diamonds) represent the numerical estimation of the mass exponent $\tau(q)$ calculated by the three
algorithms for MFA of complex networks, respectively.}\label{7}
\end{figure}

It is hard to evaluate the performance of the three algorithms
only according to the above three figures. In order to further
quantify the performance of these algorithms, we introduce the
relative standard error as in Ref. \cite{Anh2002}. Based on
the relative standard error $E$, we can determine the goodness of
the numerical results of the mass exponents $\tau(q)$ obtained
from the three algorithms compared with the analytical results for
the three deterministic model networks. The relative standard
error $E$ can be defined as
\begin{equation}
E=\frac{E_{1}}{E_{2}},
\end{equation}
where
\begin{equation}
E_{1}=\sqrt{\frac{1}{61}\sum_{q}(\tau_{t}(q)-\tau_{n}(q))^{2}},
\end{equation}
and
\begin{equation}
E_{2}=\sqrt{\frac{1}{61}\sum_{q}(\tau_{t}(q)-\bar{\tau_{t}})^{2}},
\end{equation}
the $\tau_{t}(q)$ and $\tau_{n}(q)$ represent the analytical and
numerical results of the mass exponents $\tau(q)$ respectively;
$\bar{\tau_{t}}$ is the average of the $\tau_{t}(q)$. The goodness
of fit is indicated by the result $E < 1$ \cite{Anh2002}. We
summarize the corresponding relative standard error between the
analytical and numerical results of the mass exponents $\tau(q)$
in Table I. From Table I, we see that the relative standard errors
for these three methods are all rather small. This result
indicates all these three algorithms can give correct numerical
results.

In addition, we compare the consumed CPU time of the three
algorithms for MFA of the networks generated from the three
network models. The results are given in Table II. From Table II,
 we can CBB algorithm takes a substantial amount of computation time and
memory resources, while SB algorithm consumes the least amount of
CPU time and memory resources. This is to say that the SB
algorithm has an overwhelming advantage in consuming CPU time and
memory resources. Therefore the SB algorithm can be considered as
the most effective, feasible, and accurate algorithm to calculate
the mass exponents $\tau(q)$ and then to study the multifractal
properties of complex networks among the three algorithms.

\begin{table}
\caption{The relative standard error $E$ of the three algorithms for MFA of the
networks generated from the three network models.}
\begin{tabular}{|c|c|c|c|}
\hline
     &   $(u, v)$-flower   & The minimal model &  The generalized model  \\
\hline
improved box-counting (BC) \cite{Li2014}       &  0.01406849   &  0.00533030   &  0.02067627  \\
  Compact-box-burning (CBB) \cite{Furuya2011} &  0.01567521   &  0.02449339   &  0.02245775  \\
  {\bf Sandbox (SB)}              &  0.01408235   &  0.00623890   &  0.01299720  \\
\hline
\end{tabular}
\end{table}

\begin{table}
\caption{The CPU time of the three algorithms for MFA of the
networks generated from the three network models(unit: s).}
\begin{tabular}{|c|c|c|c|}
\hline
     &   $(u, v)$-flower   & The minimal model &  The generalized model  \\
\hline
improved box-counting (BC) \cite{Li2014}       &  4069   &  8568   &  1299  \\
  Compact-box-burning (CBB) \cite{Furuya2011} &  817891   &  1003087   &  244206  \\
  {\bf Sandbox (SB)}              &  95   &  135   &  44  \\
\hline
\end{tabular}
\end{table}

\section{Applications}

Wang {\it et al.} \cite{Wang2012} applied the modified
fixed-size box-counting algorithm to study the multifractal
properties of some theoretical model networks and real networks,
such as scale-free networks, small-world networks, and random
networks. All of these complex networks have been widely used in
various studies. In this Section, we want to adopt the SB
algorithm to detect the multifractal behavior of these networks.

\subsection{Scale-free networks}

Based on the growth and preferential attachment
characteristics of many complex networks in real world,
Barab\'{a}si {\it et al.} \cite{Barabasi1999} proposed a BA model
to explain the generating mechanism of power law distributions. In
this paper, we use the BA model to generate scale-free networks
and then investigate their multifractality.

Here, we start with an initial network with $m=3$ nodes, and its
three nodes are connected each other. At each step, we add one
node to this initial network. Then this new node is connected to
$m_{0}=1$ existing nodes with probability $e$. We denote the
probability that the new node is connected to node $i$ as $e_{i}$.
The probability $e_{i}$ is defined as
$e_{i}=\frac{k_{i}}{\sum_{j}k_{j}}$, where $k_{i}$ is the degree
of node $i$.

In this paper, we respectively generated 100 scale-free networks
with 6000 nodes, 8000 nodes, and 10000 nodes. The SB algorithm is
then used to calculate their average mass exponents $\langle
\tau(q) \rangle$ and average generalized fractal dimensions
$\langle D(q) \rangle$, which are shown in Fig. 8. From Fig. 8, we find that the $\langle \tau(q) \rangle$ and $\langle D(q)
\rangle$ curves of scale-free networks are convex. So
multifractality exists in these scale-free networks.

\begin{figure}
\centerline{\epsfxsize=8.5cm \epsfbox{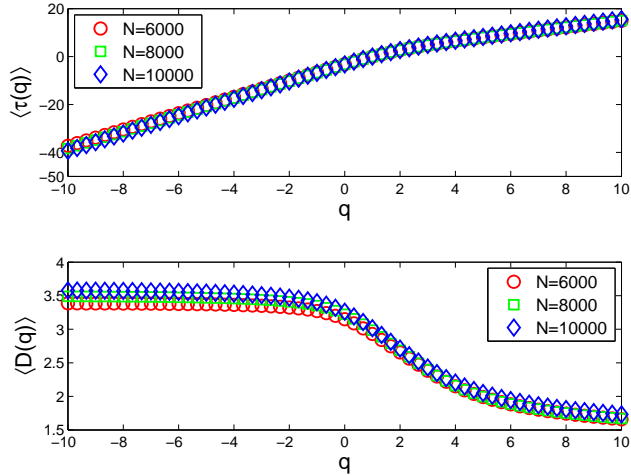}}
\caption{(Color online) Average mass exponents $\langle \tau(q) \rangle$ (upper panel) and average generalized fractal dimensions
$\langle D(q) \rangle$ (lower panel) for the scale-free networks derived from the SB algorithm. Here the average is calculated from
100 realizations.}\label{8}
\end{figure}

\subsection{Small-world networks}

Based on the random rewiring procedure, Watts and Strogatz
\cite{Watts1998} introduced the WS small-world network model which
is a transition from a completely regular network to a completely
random graph. Small-world networks not only retain the high
clustering coefficient of regular networks, but also capture the
short average path length of random graphs. Newman and Watts
\cite{Newman1999} proposed a NW model which is a modified version
of the original WS model. In the NW model, the shortcuts are added
between randomly chosen pairs of nodes, but no connections are
removed from the regular lattice.

The NW model can be described as follows. Firstly, we start with a
regular graph. We consider the nearest-neighbor coupled network
containing $N$ nodes. Each node of the coupled network is
connected to its $K$ nearest-neighbors by undirected edges, where
$K$ is an even integer. Secondly, we randomly add some new
connections to the coupled network. With probability $e$, we
connect the pair of nodes chosen randomly.

In this paper, we first generated the three nearest-neighbor
coupled networks containing 6000 nodes, 8000 nodes, and 10000
nodes, respectively. And we set $K=4$ so that each node of these
networks is connected to its 4 nearest-neighbors. Then we added a
connection between pairs of nodes of the three coupled networks
with probability $e=0.0008$, $e=0.0005$, and $e=0.001$,
respectively. For each case, we generated 100 small-world networks
using the NW model. Next, we applied the SB algorithm to calculate
their average mass exponents $\langle \tau(q) \rangle$ and average
generalized fractal dimensions $\langle D(q) \rangle$. The
$\langle \tau(q) \rangle$ and $\langle D(q) \rangle$ curves are
plotted in Fig. 9. From Fig. 9, we find that the $\langle
\tau(q) \rangle$ and $\langle D(q) \rangle$ curves of small-world
networks are not straight lines, but the fluctuation of these
curves is very small. This means that the multifractality of these
small-world networks is not obvious.

\begin{figure}
\centerline{\epsfxsize=8.5cm \epsfbox{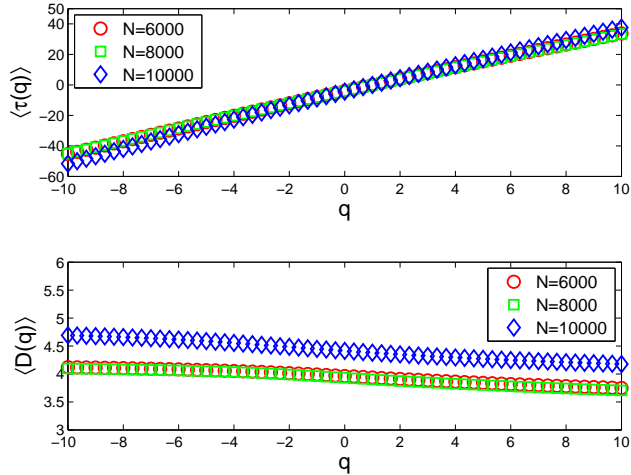}}
\caption{(Color online) Average mass exponents $\langle \tau(q) \rangle$ (upper panel) and average generalized fractal dimensions
$\langle D(q) \rangle$ (lower panel) for the small-world networks derived from the SB algorithm. Here the average is calculated
from 100 realizations.}\label{9}
\end{figure}

\subsection{Random networks}

The Erd\H{o}s-R\'{e}nyi (ER) random graph model
\cite{Erdos59} is one of the classical network models for
generating a completely random network. We start with $N$ isolated
nodes. For every possible pair of nodes, we connect them by an
undirected connection with probability $e$.

In this paper, we considered the three ER random graph models
containing 6000 nodes, 8000 nodes, and 10000 nodes. Then we
connected all possible node pairs of the three ER models with
probability $e=0.0008$, $e=0.00054$, and $e=0.0005$, respectively.
For each case, we generated 100 random networks by using the ER
model and extract the largest connected component from each random
network. Next, we used the SB algorithm to calculate the average
mass exponents $\langle \tau(q) \rangle$ and average generalized
fractal dimensions $\langle D(q) \rangle$ of these largest
connected components. In Fig. 10, we show the $\langle \tau(q)
\rangle$ and $\langle D(q) \rangle$ curves. As we can see from
Fig. 10, the $\langle \tau(q) \rangle$ and $\langle D(q) \rangle$
curves of random networks are close to straight lines. So this is
to say, the multifractality almost does not exist in these random
networks.

\begin{figure}
\centerline{\epsfxsize=8.5cm \epsfbox{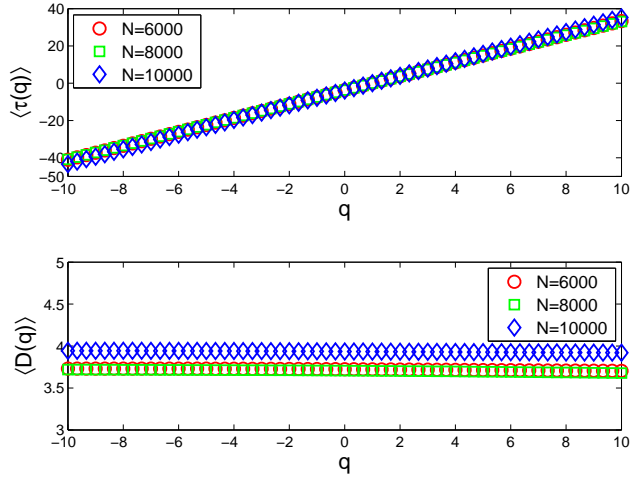}}
\caption{(Color online) Average mass exponents $\langle \tau(q) \rangle$ (upper panel) and average generalized fractal dimensions
$\langle D(q) \rangle$ (lower panel) for the random networks derived from the SB algorithm. Here the average is
calculated from 100 realizations.}\label{10}
\end{figure}

\section{Conclusion}

In this work, we employed the sandbox (SB) algorithm
proposed by T\'{e}l {\it et al.} \cite{Tel1989}, for multifractal analysis (MFA) of complex networks.

First we compared the SB algorithm with two existing algorithms of
MFA for complex networks: the compact-box-burning (CBB) algorithm
proposed by Furuya and Yakubo \cite{Furuya2011}, and the improved
box-counting (BC) algorithm proposed by Li {\it et al.}
\cite{Li2014}, by calculating the mass exponents $\tau(q)$ of some
deterministic model networks (networks generated from the $(u,
v)$-flower, the minimal model, and the generalized version of the
minimal model). We made a detailed comparison between the
numerical results and the theoretical ones of these model
networks. The comparison results show that the SB algorithm is the
most effective, feasible, and accurate algorithm to calculate the mass
exponents $\tau(q)$ and to explore the multifractal behavior of
complex networks.

 In addition, we applied the SB algorithm to study the multifractality of some
classic model networks, such as scale-free networks, small-world
networks, and random networks. Our numerical results show that
multifractality exists in scale-free networks, that of small-world
networks is not obvious, and it almost does not exist in random
networks.

\section*{Acknowledgments}

This project was supported by the Natural Science Foundation of
China (Grant No. 11371016), the Chinese Program for Changjiang
Scholars and Innovative Research Team in University (PCSIRT)
(Grant No. IRT1179), the Lotus Scholars Program of Hunan province
of China.

\end{document}